\documentclass[doublecol]{epl2} 
\usepackage{graphicx,amsfonts,amssymb,amsmath, hyperref}
\usepackage[applemac]{inputenc}
\usepackage{ifpdf}

\newif\ifhyper
\hypertrue
\ifhyper
\hypersetup{
  citecolor = {green},
  colorlinks = {true}, 
  urlcolor = {blue} 
}
\fi

\newlength{\ldag}
\def\nbR{\ensuremath{\mathrm{I\!R}}} 

\title{Nonperturbative renormalization group approach to Lifshitz critical behaviour}

\author{K. Essafi\inst{1} \and J.-P.  Kownacki\inst{2} \and D. Mouhanna\inst{1}}
\shortauthor{F. Author \etal}

\institute{\inst{1} LPTMC,
CNRS UMR 7600, UPMC, 4 Place Jussieu, 75252 Paris Cedex 05, France\\
  \inst{2} LPTM, CNRS UMR 8089-Universit\'e de Cergy-Pontoise,   2 avenue Adolphe Chauvin, 95302 Cergy-Pontoise  Cedex, France}


\abstract{The  behaviour of a $d$-dimensional vectorial $N=3$    model  at  a $m$-axial Lifshitz critical point is investigated by means of a nonperturbative renormalization group approach  that   is  free of the huge  technical difficulties  that plague  the perturbative approaches  and   limit   their  computations    to the lowest orders.   In particular  being systematically  improvable,  our approach  allows  us  to  control  the  convergence of  successive approximations  and thus to get reliable physical  quantities in $d=3$.}

\begin{document}

\maketitle

\section{Introduction}

Lifshitz critical behaviour  (LCB) \cite{hornreich75} (see also  \cite{hornreich80,selke92,diehl02,henkel10})   occurs   when   a disordered phase  encounters 
 both a homogeneous  ordered phase and a spatially modulated ordered phase with a modulation wave-vector ${\bf q}_{\tiny\hbox{mod}} \ne \bf 0$.   In the general
  case the  vector ${\bf q}_{\tiny\hbox{mod}} $  spans a $m$-dimensional subspace of the $d$-dimensional space with $0\le m \le d$.    For  a $N$-component order 
   parameter  the universal behaviour at criticality is completely determined by the set    $(m,d,N)$.   LCB   has been proposed  to occur in many
    systems including magnetic models  (notably the ANNNI model \cite{selke88}),  liquid crystals,  microemulsions, polymer mixtures, ferroelectrics, 
    high-$T_c$ superconductors, see  \cite{diehl02,henkel10}  for reviews.  In the domain of magnetic materials there has  been  a growing activity in the search 
    for  LCB behaviour. A clear-cut  LCB has been found in manganese  phosphide  (MnP)  \cite{becerra80}  and,  possibly, in the ternary
     uranium silicide (UPD$_2$Si$_2$)  \cite{plackowski11}.  One can thus expect accurate  determinations of the  critical quantities from experiments 
      in a near future. 

From the theoretical point of view the simplest model displaying  LCB can be obtained by generalizing the Hamiltonian, or action, relevant to study
 the usual vectorial  ferromagnetic-paramagnetic  phase transition. Let us consider a  $N$-component  vector field  ${\boldsymbol{\phi}}({\boldsymbol{x}})$ 
  in a $d$-dimensional space.  The coordinates ${\boldsymbol x}$ are decomposed  into   a {\it parallel}  component  $\boldsymbol{x}_{\parallel}\in \nbR^m$ and an {\it orthogonal}  component ${\boldsymbol  x}_{\perp}\in \nbR^{d-m}$, {\it i.e.}  ${\boldsymbol x}=({\boldsymbol x_{\parallel}},{\boldsymbol x_{\perp}})$.   The   action allowing a LCB  reads: 
\begin{equation}
\begin{array}{ll}
\Gamma[{\boldsymbol{\phi}}] = \displaystyle \int \text{d}^{d-m}x_{\perp} \text{d}^{m}x_{\parallel}&\Bigg\{\displaystyle {Z_{\parallel} \over 2}(\boldsymbol{\partial}_{\parallel}^2{\boldsymbol{\phi}})^2 + 
{Z_{\perp} \over 2}(\boldsymbol{\partial}_{\perp}{\boldsymbol{\phi}})^2 \\ & \displaystyle \hspace{-0.5cm} + {\rho_0 \over 2}(\boldsymbol{\partial}_{\parallel}{\boldsymbol{\phi}})^2 
\displaystyle +\,  u \left({{{\boldsymbol{\phi}}^2}\over 2}-\kappa\right)^2 \Bigg\}
\label{effectiveaction}
 \end{array}
\end{equation}
where $\boldsymbol{\partial_{\parallel}}$ and $\boldsymbol{\partial_{\perp}}$ stand for the derivatives in the corresponding directions.  The coupling constants $Z_{\parallel}$,  
 $Z_{\perp}$  and $u$ are  supposed to be always positive while   $\rho_0$  and $\kappa$ are  allowed to change sign.  The coupling $\kappa$  stands for a magnetization 
 occuring in the -- homogeneous  -- ordered phase.  From a  mean-field analysis one observes that, for  $\rho_0>0$,  when the  coefficient  $\tau=-u\kappa$
  in front of  ${\boldsymbol{\phi}}^2$  varies from a positive  to a negative value the  system undergoes a phase transition  from  a 
  disordered  to a {\it homogeneous} ordered phase while  for   $\rho_0<0$ a transition  occurs,  for some  $\tau=\tau_c$,   from a  disordered to a 
  {\it modulated}  ordered phase.  The two transition lines join at the Lifshitz point which, within  a mean-field  analysis,  is located at  $\tau=\rho_0=0$.

The  salient property characterizing  LCB is that  of {\it anisotropic scale invariance} (ASI).  Indeed,  at the Lifshitz point, 
because of the absence of  $(\boldsymbol{\partial}_{\perp}^2{\boldsymbol{\phi}})^2$  term,   the scaling dimensions  in the $\perp$  and 
$\parallel$ directions differ.  In particular   the two-point correlation functions  scale as \cite{diehl02,henkel10}
\begin{align}
\Gamma^{(2)} ({\bf q}_\perp \rightarrow {\bf 0}, 
{\bf q}_\parallel = {\bf 0}) &\sim {\bf q}_\perp^{2-\eta_{\ell 2}}  \nonumber  \\
\Gamma^{(2)} ({\bf q}_\perp = {\bf 0}, {\bf q}_\parallel \rightarrow {\bf 0}) &\sim {\bf q}_\parallel^{4-\eta_{\ell 4}} \ , 
\label{scaling}
\end{align}
which define the two  anomalous scaling dimensions $\eta_{\ell 2}$ and $\eta_{\ell 4}$.  On the other hand,   for  a generic scaling operator,  one expects   the following asymptotic behaviour  under a scale transformation: ${\cal O}(s\,  
  {\bf q}_{\perp},s^{\theta}\,  {\bf q}_{\parallel}) \sim s^{-\Delta} {\cal O}({\bf q}_{\perp}, {\bf q}_{\parallel})$ when ${s\to 0}$,  where $\Delta$ is  the 
  scaling dimension associated to the operator ${\cal O}$,  $\theta$ being the anisotropy critical exponent. In particular,  for the two-point  function one has: 
  $\Gamma^{(2)} (s {\bf q}_\perp, s^{\theta} {\bf q}_\parallel )\sim   s^{2-{\eta_{\ell 2}}} \Gamma^{(2)}({\bf q}_\perp, {\bf q}_\parallel)$ when ${s\to 0}$. This
   behaviour, together with Eq.(\ref{scaling}),   provides the relation: 
   \begin{equation}
 {\theta}  = {2 -{\eta_{\ell 2}}\over 4 - {\eta_{\ell 4}}}\ . 
   \label{z}
   \end{equation}
 Finally   two critical exponents, ${\nu_{\ell 4}}$ and  ${\nu_{\ell 2}}$,  characterize the behaviour  of  the correlation lengths near criticality: 
 
 \begin{equation}
 \noindent \xi_{\parallel}\propto \tau^{-{\nu_{\ell 4}}}  \hspace{0.5cm} {\hbox{and}}  \hspace{0.5cm} \xi_{\perp}\propto \tau^{-{\nu_{\ell 2}}} 
 \nonumber
  \end{equation}
 
 with

  \begin{equation}
{\nu_{\ell 4}}={\theta}\, {\nu_{\ell 2}}\ .
\label{nunu}
 \end{equation}

ASI  occurs  in many contexts:  equilibrium critical phenomena of anisotropic systems, like  those described by action (\ref{effectiveaction})  at a  Lifshitz point 
 or, {\it e.g.}  in the crumple-to-tubule transition in anisotropic membranes \cite{radzihovsky95,essafi11}, as well as in dynamical critical phenomena at
  and away from equilibrium (see \cite{diehl02}).  In quantum  field theory an intensive activity has  been developed  towards  theories for which  Lorentz 
  invariance is broken at high energy by high order derivative terms  in the spatial directions (see  \cite{alexandre11} for a review). In these 
  ``Lifshitz-type  theories"   the presence  of   anisotropy between temporal and spatial directions  drastically improve the UV behaviour and
   renormalizability properties.   These ideas have been  further  extended  towards  anisotropic scale  invariant gravity   \cite{horava09}  and  cosmology   \cite{nakayama11}.  Finally  a theory of local  scale invariance (LSI) has been  introduced   \cite{henkel02} both for equilibrium and out of equilibrium  phenomena 
   leading  to  conjecture exact   expressions for  the two-point  correlators of anisotropic systems.   While Monte Carlo results \cite{pleimling01} have been
    claimed to agree with  these predictions,  in a very recent work  \cite{rutkevich11}  the predictions of the LSI theory of  \cite{henkel02}    were challenged. Specifically, reference  \cite{rutkevich11}  found that the epsilon expansions of some scaling functions obtained from a two-loop expansion about the upper critical dimension are inconsistent with the predictions of  \cite{henkel02}   and  \cite{pleimling01}.

\section{Nonperturbative renormalization group approach}

In this  context it  is clear  that  an efficient and systematically improvable  approach  of anisotropic systems, and in particular of LCB,  is  needed. From this point of view  one has to emphasize  that the available, perturbative,  techniques   are especially in trouble.   Let us  start  with  the weak-coupling $\epsilon$-expansion. A first problem is that  going from isotropic to anisotropic systems shifts the upper critical dimension  from $d_{uc}=4$  to $d_{uc}=4+m/2$.  This means that even for the minimal non trivial value  of $m$, equal
 to one, the $\epsilon$-expansion implies to deal with the  large value $\epsilon=3/2$  when   computing  the critical properties in $d=3$.   Assuming 
 that the series obtained are Borel-summable, which  is not guaranteed,   getting  reliable physical quantities thus  implies   computing, at least  up to four or five-loop order.  But then  
 one faces a second and important problem. As emphasized in \cite{diehl00, shpot01,diehl02} the real-space  free propagator  used to perform the  
 $\epsilon$-expansion  takes a very complicated form,  known as Fox-Wright generalized hypergeometric functions, that leads to  {\sl enormous}   \cite{shpot08} technical difficulties. This explains why  the weak-coupling $\epsilon$-expansion results  have been very controversial  during a long time  \cite{shpot01,diehl01} and  that it took 
  almost  twenty  years to fill the gap between early  one-loop order results  \cite{hornreich75} and the complete two-loop order  computation 
  \cite{mergulhao,diehl00, shpot01,diehl02}.  For this reason  it is  extremely  unlikely that higher-order contributions  will be obtained in a near future. Similar 
  difficulties  occur  within a large-$N$  approach (see \cite{shpot05} and \cite{shpot12})  and  it is only very recently \cite{shpot08} that consistency between 
  this large-$N$ approach at order $O(1/N)$ and the weak-coupling expansion at two-loop order has been firmly established.  Finally  note that it is also possible, in 
  principle,  to investigate LCB  by means of  a low-temperature approach in the vicinity of the lower  critical 
  dimension which, for  $N>1$ components system, is given by $d_{lc}=2+m/2$.  This has been done at one-loop order by Sak and Grest \cite{grest78}. However, 
  as in the  $O(N)$ model,  the series  obtained  within a low-temperature approach are generally suspected to be non-Borel-summable and  
  thus of no practical use.

We investigate here  the  LCB   by means of  a nonperturbative renormalization group (NPRG) approach.  Our computation is based  on the concept of  
 running effective action  \cite{wetterich93c}  (see \cite{bagnuls01,berges02,delamotte03,pawlowski07,rosten12}  for reviews),  $\Gamma_k[\boldsymbol \phi]$,   a 
 functional of   the $N$-component  vector field $\boldsymbol \phi({\bf x})$ that describes the effective physics  at a  coarse grained 
 scale $k$. Technically   the index  $k$  stands for  a running scale that separates the high-momentum modes,  with $q>k$,  from  the low-momentum ones, with $q<k$ and   
 $\Gamma_k[\boldsymbol \phi]$  represents   a coarse grained free energy where only fluctuations with momenta $q\ge k$  have been integrated out.  The  
 running  of $k$ towards   $k=0$ thus corresponds to  gradually integrate over all fluctuations. The $k$-dependence, RG flow, 
  of $\Gamma_k$ is  provided  by an exact -- albeit one-loop -- evolution equation \cite{wetterich93c}:
\begin{equation}
{\partial \Gamma_k\over \partial t}={1\over 2} \hbox{Tr} \left\{(\Gamma_k^{(2)}+R_k)^{-1}
 {\partial R_k\over \partial t}\right\}
\label{renorm}
\end{equation}
where $t=\ln \displaystyle {k / \Lambda}$, $\Lambda$ being some microscopic, lattice, scale. The trace in (\ref{renorm}) involves  a 
 $d$-dimensional  momentum  integral over a momentum $\bf q$ as well as a summation over vectorial  indices. The function $R_k(q)$  
 realizes  the split between low- and high-momentum degrees of freedom while   $\Gamma_k^{(2)}$ represents  the second functional derivative  of  $\Gamma_k$ with respect to $\boldsymbol \phi$, {\it i.e.} the inverse field-dependent propagator.  Considering  $\Gamma_k$  in its full generality  Eq.(\ref{renorm})  provides  an {\it exact}  
  RG flow for  the coupling constants associated to any power of $\boldsymbol \phi$ and of its derivatives.  
  
   There are several  major  advantages in using   Eq.(\ref{renorm}).   First,   one deals with {an}  one-loop  equation while the   computations are naturally performed  in   momentum  space.  In this way we avoid {\it all}  the technical difficulties  encountered within the perturbative approaches   that  are associated  to  the  multi-loop structure and  the complexity of the propagator in real space. Second, the equation being nonperturbative in the coupling constants entering in the effective action (the $\phi^4$-like coupling  constant $u$, the temperature $T\sim 1/ \kappa$  as  the parameter $1/N$)  the approach  overcomes   a major  problem of the perturbative  theory: the need to  resum,  if  possible,   the perturbative renormalized series.   Third  this technique is systematically  improvable without  conceptual or  technical difficulty. Let us develop these  last  two points.     Eq.(\ref{renorm}), although exact is not  exactly solvable. One thus has to  consider truncations of  $\Gamma_k[\boldsymbol \phi]$ and, thus,  {\it approximations} of Eq.(\ref{renorm}).  Different kinds of approximations are allowed  which  keep the nonperturbative  and one-loop character  of the equation untouched.  A very useful and efficient approximation  is based on an expansion of 
  $\Gamma_k[\boldsymbol \phi]$  in powers of both  fields and field derivatives. A derivative  expansion is particularly  justified  to 
  investigate critical  phenomena whose physics  is dominated by  low momenta and thus by  low  powers of the field derivatives.   A  field  expansion is  justified  by both its general character and  its rapid convergence  \cite{canet03a}   apparently  without  need  of resummation procedure, as attested by several studies involving  Ising model  {\cite{canet03b}, frustrated   magnets \cite{tissier00,delamotte03}, randomly dilute Ising model \cite{tissier01b},  membranes \cite{kownacki09,essafi11} and  other \cite{berges02}.  
Moreover in  the context of anisotropic systems  a field expansion is particularly suitable since the physical dimension 3  is close to  the {\it lower } critical dimension  of (vectorial) anisotropic systems  $d_{lc}=2+m/2$  in the vicinity of which,  the RG  flow (\ref{renorm}),  together with the ansatz (\ref{effectiveaction}),   is one-loop exact.   
One can thus expect  the ansatz  (\ref{effectiveaction}), or  a  bit more sophisticated ansatz,   to provide   very sensible  results in  $d=3$.   In this article we provide the RG  equations for the coupling constants entering in the action  (\ref{effectiveaction}) while we  have  computed with powers of the field up to order $\boldsymbol{\phi}^{12}$.  The validity of this  approach is  then  checked by  studying  both   the cut-off independence of the physical quantities and  their behaviour  when  the field content is enriched.  We show in particular that  converged  critical quantities are obtained  using   a  limited number of  powers of the field.  Note  that a NPRG approach of  LCB in the Ising case  has already been 
performed in \cite{bervillier04} using a  full  but local potential approach of the Polchinski equation,  thus neglecting the anomalous dimension. We investigate here the behaviour of the vectorial $N=3$ case  providing  both   the critical exponents ${\nu_{\ell 4}}$ and ${\nu_{\ell 2}}$  together with   the anomalous scaling dimensions  ${\eta_{\ell 4}}$ and  ${\eta_{\ell 2}}$.

\section{Renormalization group equations}

The flow equations for the coupling constants $\kappa$, $u$, $\rho_0$ entering in  (\ref{effectiveaction}) are obtained, as usual \cite{berges02,delamotte03} by appropriate functional 
derivatives of the RG equation (\ref{renorm}).  One defines  the dimensionless quantities using the scale $k_{\perp}$\footnote{The scale  
 $k_{\parallel}=k_{\perp}^{\theta}$  could be chosen as well.}: 
 $\overline \kappa=Z_{\perp}^{(4-m)/4} Z_{\parallel}^{m/ 4} k_{\perp}^{(m+4-2d)/2}\kappa$, $\overline  u =Z_{\perp}^{(m-8)/ 4} 
 Z_{\parallel}^{-m/4} k_{\perp}^{(2d-m-8)/2} u$ and $\overline \rho_0=Z_{\perp}^{-1/2} Z_{\parallel}^{-1/2}  k_{\perp}^{-1}\rho_0$ and their flow  reads, with $t=$ln $k_{\perp}/\Lambda$:
\begin{equation}
\begin{array}{ll}
\partial_t \overline \kappa   = & \displaystyle - \left(d - m + {\theta}(m + {\eta_{\ell 4}} - 4)\right) \overline \kappa  \\
& \displaystyle  + (N-1) \overline l_{\perp,2}^0 + 3 \overline l_{\parallel,2}^0 \\
\\
\partial_t \overline u  = & \displaystyle \left(d - m + {\theta}(m + 2{\eta_{\ell 4}} - 8) \right) \overline u \\
& \displaystyle  + 2 \overline u^2 \left((N-1) \overline l_{\perp,4}^0+ 9 \overline l_{\parallel,4}^0\right)\\
\\
\label{couplingsflow}
\displaystyle \partial_t \overline \rho_0  =  &  \displaystyle {\theta}({\eta_{\ell 4}}-2) \overline \rho_0 + \frac{1}{m}   \left({1\over  \overline u\  \overline\kappa^2} \left(\overline M_{\perp,2}^1- \overline M_{\parallel,2}^1\right) \right. \\
&  \displaystyle  \left.  - 
{2\over \overline\kappa} \left(\overline M_{\perp,4}^1 - \overline M_{\parallel,4}^1\right)  \right)\ , 
\end{array} 
\end{equation}
while the  {\it running} anomalous dimensions ${\eta_{\ell 2}=- \partial_t \ln Z_{\perp}}$ and ${\eta_{\ell 4}=- {(1/\theta)}\, \partial_t \ln Z_{\parallel}}$ are given by:
\begin{equation}
\begin{array}{ll}
{\eta_{\ell 2}} = &\displaystyle \frac{1}{\overline \kappa}  \left(\overline l_{\perp,2}^0 + \overline l_{\parallel, 2}^0 \right) - 
\frac{1}{2 \overline u ~ \overline \kappa^2}  \left(\overline l_{\perp, 0}^0 - \overline l_{\parallel, 0}^0\right)\ , \\
\\
{\eta_{\ell 4}}    = & \hspace{-0.3cm}\displaystyle \frac{1}{6{\theta}m (m+2)} \frac{1}{\overline u^3 \overline \kappa^4}  \left((m+2) \overline u^2 \overline \kappa^2 
 \left(\overline S_{\perp,4}^{1} - \overline S_{\parallel,4}^1 \right) \right. \\
 & \hspace{-0.3cm} \displaystyle  - 2(m+2) \overline u^3 \overline \kappa^3  \left(\overline S_{\perp,6}^1 + \overline S_{\parallel,6}^1 \right) - \frac{9}{2} 
 \left(\overline T_{\perp,2}^2- \overline T_{\parallel,2}^2 \right)  
\\
& \displaystyle  \hspace{-0.3cm} + 9 \overline u\, \overline \kappa  \left(\overline T_{\perp,4}^2 + \overline T_{\parallel,4}^2 \right) - 8 \overline u^2\,  \overline
 \kappa^2   \left(\overline T_{\perp,6}^2 - \overline T_{\parallel,6}^2 \right)  \\
& \displaystyle  \hspace{-0.3cm} + 4 \overline u^3\,  \overline \kappa^3 \left(\overline T_{\perp,8}^2+ \overline T_{\parallel,8}^2 \right) + 6 \overline u^2 \overline\kappa^2  
\left(\overline U_{\perp,2}^2  - \overline U_{\parallel,2}^2\right) \\
&\displaystyle  \hspace{-0.3cm} \left.- 12 \overline u^3\,  \overline \kappa^3 \left(\overline U_{\perp,4}^2 + \overline U_{\parallel,4}^2\right) \right)\ , 
\label{eta}  
\end{array} 
\end{equation}
where, in Eqs.(\ref{couplingsflow}) and   (\ref{eta}),  $\overline  l_{a, b}^{\alpha}$, $\overline M_{a, b}^{\alpha}$, $\overline S_{a, b}^{\alpha}$, 
$\overline T_{a, b}^{\alpha}$, $\overline   U_{a, b}^{\alpha}$ are  dimensionless  "threshold functions"  (see \cite{berges02})  $l_{a, b}^{\alpha}$, 
$M_{a, b}^{\alpha}$, $S_{a, b}^{\alpha}$, $T_{a, b}^{\alpha}$, $U_{a, b}^{\alpha}$   that are  given by :
\begin{equation}
  {\cal A}_{a, b}^{\alpha} =   \displaystyle  \widehat{\partial \over \partial t} \int \text{d}^m q_{\parallel}\,   { K_{b}\   
  {\bf q}_{\parallel}^{2\alpha}\  F({\bf q}_{\parallel})\over \left[P({\bf q}_{\parallel})+  m_{a}^2\,  {\bf q}_{\parallel}^2\right]^{\gamma_b}}
\label{threshold}
\end{equation}
where $K_{b}=-\Gamma[\gamma_{b}](4\pi)^{(m-d)/2} / 2 $,  $\gamma_b = (m+b-d)/2$, $P({\bf q}_{\parallel})=Z_{\parallel}\, 
{\bf q}_{\parallel}^4+\rho_0 {\bf q}_{\parallel}^2+R_{k_{\perp}}({\bf q}_{\parallel})$,   $m_{\parallel}^2=4  u  \kappa$, $m_{\perp}=0$ and 
where $\widehat{{\partial}}/{\partial t}$ only acts on     
$R_{k_{\perp}}$. In Eq.(\ref{threshold})  the function  $F({\bf q}_{\parallel})$ is given by  $1$, $(dP/d{\bf q}_{\parallel}^2)^2$, 
 $(dP/d{\bf q}_{\parallel}^2)^3$,  $(dP/d{\bf q}_{\parallel}^2)^4$ and  $(d^2P/d({\bf q}_{\parallel}^2)^2)^2$ for $l$, $M$, $S$, $T$ and $U$ respectively.  
 These threshold functions encode the nonperturbative content of the approach since, as it is clear from (\ref{threshold}),  they are nonpolynomial 
 functions of  the coupling constants $u$ and $\kappa$ entering in the squared  "mass"  $m_{\parallel}^2$.     Note that   the threshold functions ${\cal A}_{a, b}^{\alpha}$  of Eq.(\ref{threshold}) are  integrals  over  ${\bf q}_{\parallel}$ only. Indeed  since the momenta   ${\bf q}_{\perp}$, or the derivative $\boldsymbol{\partial}_{\perp}$,  enters only quadratically in action (\ref{effectiveaction})  one can exactly perform the integration over  the  $d-m$ orthogonal  degrees of freedom in the  {RG equations (\ref{couplingsflow}) and   (\ref{eta})}.

\section{Physical results}

Let us discuss the RG equations Eqs.(\ref{couplingsflow}) and   (\ref{eta}). The  one-loop structure of  Eq.(\ref{renorm})  together  with 
action (\ref{effectiveaction})  allow, when  the Eqs.(\ref{couplingsflow}) and   (\ref{eta}) are expanded in powers of the suitable coupling constant,  to recover  all the  one-loop results obtained perturbatively.  The weak-coupling results obtained in the vicinity of the upper critical dimension 
\cite{hornreich75} are easily recovered  by  performing an expansion in powers of  the coupling constant $u$ in the vicinity of $d_{uc}=4+m/2$.  
In the same way one obtains the large-$N$ results at dominant order \cite{hornreich75b}.  More importantly for our purpose we recover the low-temperature
 $T$ results obtained in the vicinity of the lower critical dimension $d_{lc}(m)=2+m/2$ by  Sak and Grest \cite{grest78}  using a large-$\overline\kappa$ 
 expansion since $\kappa\sim 1/T$. We get  the flow of $\overline \kappa$: 
  $\partial_t \overline \kappa= -2 \epsilon \overline \kappa + \tilde C(N-2)$ with $\tilde C=\Gamma[m/4]/2^d \pi^{d/2} \Gamma[m/2]$  and $\epsilon=d-d_{lc}(m)$ with the  
  running anomalous dimensions : ${\eta_{\ell 4}}=\tilde C/\overline  \kappa$ and ${\eta_{\ell 2}}=\tilde C {\theta}/\overline \kappa$. At the fixed point one has 
  $\overline \kappa^*=\tilde C(N-2)/2\epsilon$ and ${\eta_{\ell 4}}^*= 2 \epsilon/(N-2)$ and ${\eta_{\ell 2}}^*={\eta_{\ell 4}}^*/2$ which coincide exactly with 
  the expressions of Sak and Grest.  While  fully expected   this result  is particularly valuable for anisotropic systems for which the  lower 
  critical dimension $d_{lc}(m=1)=2.5$ is especially close to the physical dimension $d=3$.

  We now specialize to  the $N=3$ case, and thus, $m=1$.  One finds  a nontrivial fixed point  with two directions of instability -- corresponding to  LCB  -- in any  dimension between   $d_{lc}=2.5$ and $d_{uc}=4.5$.  Fig.\ref{etadim} displays the curves   ${\eta_{\ell 4}}$  and  ${\eta_{\ell 2}}$ as functions of $d$, that  call for several remarks.  First they show the ability of the NPRG approach to  interpolate smoothly between $d_{lc}$ and $d_{uc}$. {Second they   confirm, by a direct investigation in $d=3$ and for $N=3$,  the salient fact that   LCB  in $d=3$ is characterized by a {\sl negative}  value of ${\eta_{\ell 4}}$}.  This result   also points out the limits of the perturbative,    large-$N$ or low-temperature,  approaches  that lead to a positive value of   ${\eta_{\ell 4}}$ in $d=3$.   We now focus on the $d=3$ case.  For each fixed point  the critical exponents ${\nu_{\ell 4}}$, ${\nu_{\ell 2}}$,   ${\eta_{\ell 4}}$ and  ${\eta_{\ell 2}}$ are computed and {\it optimized} \cite{canet03a,litim02}.  
  
\begin{figure}[htbp]
\begin{center}{\includegraphics[width=0.4\textwidth]{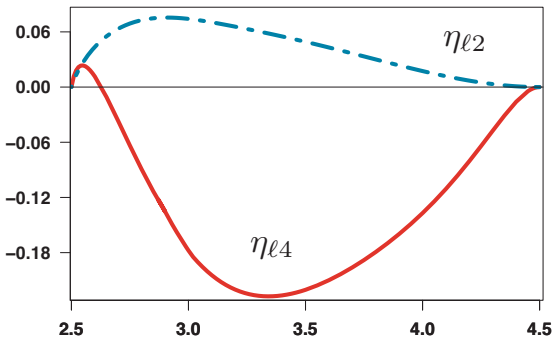}}
\end{center}
\caption{The anomalous dimensions ${\eta_{\ell 4}}$ and  ${\eta_{\ell 2}}$  as functions of the dimension $d$ using a field truncation up to $\phi^8$. }
\label{etadim}
\end{figure}

To do this one considers  one (or more)  family  of cut-off  functions indexed by a real parameter $\lambda$:  $R_{k_{\perp}}^{\lambda}({\bf q}_{\parallel})$.  Typically one has considered  a cut-off function of the form $R_{k_{\perp}}^{\lambda}({\bf q}_{\parallel})=\lambda\, Z_{\parallel} /  (\hbox{exp}({\bf q}_{\parallel}^4/k_{\perp}^{4{\theta}}-1)$. For each family one varies  $\lambda$ in order  to  find stationary values  of  the critical quantities. Stationarity is  a condition that must necessarily  be fulfilled by any  putative physical quantity  to ensure its  quasi-independence with respect to both the cut-off function and  the truncation used  \cite{canet03a}.  However an explicit study of the convergence is necessary to get  trustable results.  This has been  realized by adding  successively   powers of the field up to order $\boldsymbol{\phi}^{12}$.  Doing  this we have been able,  for  all critical exponents,  at almost any order of the field expansion\footnote{The $\phi^6$ case seems to be  special in the sense that it does not exhibit clear stationary values.}   to  find  stationary values. This is illustrated, for instance,  in Figs.\ref{fignu} and \ref{figeta}  which represent the critical exponents ${\nu_{\ell 2}}$ and ${\eta_{\ell 2}}$ in the vicinity of their stationary values for different truncations of the action.

\begin{figure}[htbp]
\begin{center}{\includegraphics[width=0.45\textwidth]{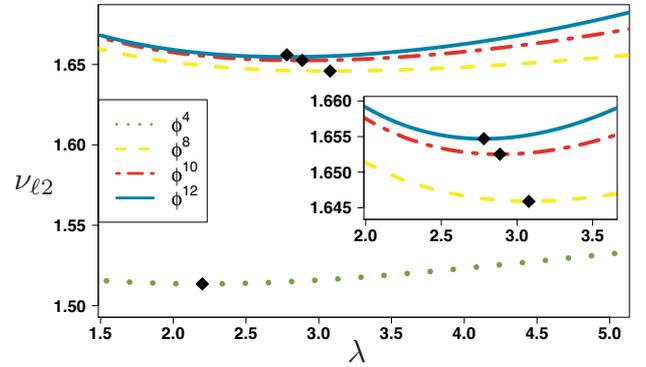}}
\end{center}
\caption{The exponent ${\nu_{\ell 2}}$ as function of  $\lambda$ for truncations  from $\phi^4$ (lower curve)  to $\phi^{12}$ (upper curve). Stationary points  are indicated by  black diamonds.}
\label{fignu}
\end{figure}

\begin{figure}[htbp]
\begin{center}{\includegraphics[width=0.45\textwidth]{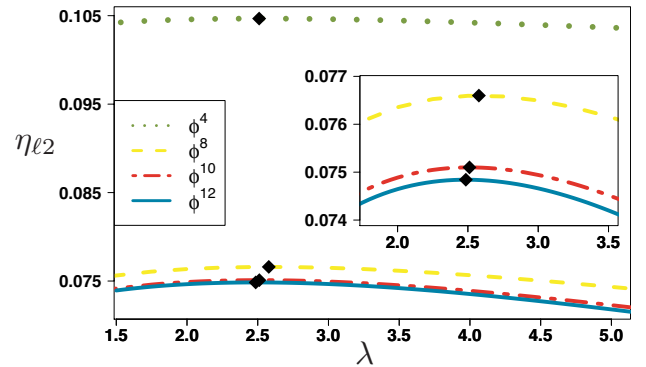}}
\end{center}
\caption{The anomalous dimension ${\eta_{\ell 2}}$  as function of   $\lambda$ for truncations from $\phi^4$  (upper  curve)  to $\phi^{12}$ (lower curve). Stationary points  are indicated by  black diamonds.}
\label{figeta}
\end{figure}

    Note   that  the critical exponents  vary  very smoothly with the parameter $\lambda$ around the stationary points  which indicates a very  weak dependence of the results with respect  to the cut-off function used.  This fact has been confirmed by using other families of cut-off functions  that lead to the same results.  More importantly Figs.\ref{fignu} and \ref{figeta} also indicate  a rapid convergence of the  physical quantities when powers of the field $\phi$  are added.  Between order $\boldsymbol{\phi}^{10}$ and  order  $\boldsymbol{\phi}^{12}$ only the third digit of ${\nu_{\ell 2}}$ and   ${\eta_{\ell 2}}$  are modified (Note however that the parallel exponents fluctuate a little bit more).  This  fact  indicates that  our  results are well converged.   They are summarized in  Table (\ref{table1}) in the column NPRG, together with the weak-coupling  results at order $\epsilon^2$   \cite{shpot01}  and the large-$N$  results at order $1/N$   \cite{shpot05,shpot12}.  {Note that the error bars  on the values of  the critical exponents  are evaluated  {\it i)}  from  the direct  analysis  of the  convergence of the field expansion for a given critical exponent  when more and more powers of the fields are added {\it ii)}  from  the discrepency that is observed between the values of a critical exponent  according to the fact that it  is obtained through a direct optimization  or if  it  is obtained through  optimization  of  the combination of the critical exponents  leading to  it through the  scaling relation (\ref{z}) and (\ref{nunu}).} An important  outcome of our approach is that  the critical exponents  found here strongly differ  from those obtained within the weak-coupling perturbative approach \cite{shpot01}.   This  discrepancy is  not surprising  as  the perturbative results have been obtained  only at low orders and   yet our computation, although based on  a different  kind of approximation, shows the importance of taking account of several orders  to reach converged results.   Finally,  we note that, amazingly,  our critical  exponents  $\nu_{\ell 4}$ and $\nu_{\ell 2}$  are  close to the values computed  within  a  very recent  large-$N$ expansion  \cite{shpot12},  contrary  to $\eta_{\ell 4}$ and $\eta_{\ell 2}$.
 \begin{table}[h]
 \centering
\begin{tabular}{| c  | c | c  | c |}
\cline{1-4}
&NPRG  &O($\epsilon^2$) \cite{shpot01} & $O(1/N)$ \cite{shpot12,shpot05}\\
\hline
${\nu_{\ell 4}}$&0.78(1) &0.392  & 0.755 \\
${\nu_{\ell 2}}$&1.655(5)&0.798   & 1.575 \\
${\eta_{\ell 4}}$ &-0.18(2)&-0.021  & 0.074 \\
${\eta_{\ell 2}}$ & 0.075(1) &0.044 & 0.102 \\
\hline
\end{tabular}
\caption{Critical exponents  $\nu_{\ell 4}$, $\nu_{\ell 2}$, ${\eta_{\ell 4}}$ and ${\eta_{\ell 2}}$ for $N=3$ and $m=1$.}
\label{table1}
\end{table}} 

\section{Conclusion}

We have shown that the NPRG   approach  provides  the  critical exponents of the Lifshitz point  while  avoiding   all  the drastic difficulties encountered using perturbative approaches.    Moreover  our approach is  systematically improvable, as explicitly shown  through the present study  of the  convergence of the field expansion. Although they should  play  a less important role,   higher derivative terms can   be treated along the sames lines. Finally our work could stimulate,  in particular,  extensive  numerical  works    as well as  investigation of high-quality magnetic compound  to confirm the adequacy of our quantitative predictions.

\acknowledgements
  K. E. and D. M. are grateful to B. Delamotte  for useful discussions.


\begin{thebibliography}{10}
\expandafter\ifx\csname url\endcsname\relax\def\url#1{\texttt{#1}}\fi

\bibitem{hornreich75}
\Name{Hornreich R. M., Luban M. \and Shtrikman S.} \REVIEW{Phys. Rev.
  Lett.}{35}{1975}{1678}.

\bibitem{hornreich80}
\Name{Hornreich R. M.} \REVIEW{J. Magn. Magn. Mater.}{15-18}{1980}{387}.

\bibitem{selke92}
\Name{Selke W.} in \Book{Phase Transitions and Critical Phenomena} edited by  {C. Domb and J.
  L. Lebowitz} Edition Vol.~15 (Academic, London) 1992 p.~1.

\bibitem{diehl02}
\Name{{Diehl H.W.}} \REVIEW{Acta Phys. slovaca}{52}{2002}{271}.

{\bibitem{henkel10}
  \Name{Henkel  M. \and Pleimling M.}
  \Book{Non-equilibrium Phase Transition,  Vol} {\bf 2}:  \Book{Ageing and dynamical scaling far from equilibrium}
  \Editor{Springer}
  \Year{(Heidelberg 2010)}.}


\bibitem{selke88}
\Name{Selke W.} \REVIEW{Phys. Rep.}{170}{1988}{213}.

\bibitem{becerra80}
\Name{{Becerra C. C., Shapira Y.,  Oliveira, Jr. N. F. \and Chang T. S.}}
  \REVIEW{Phys. Rev. Lett.}{44}{1980}{1692}.

\bibitem{plackowski11}
\Name{Plackowski T., Kaczorowski D. \and Sznajd J.} \REVIEW{Phys. Rev.
  B}{83}{2011}{174443}.

\bibitem{radzihovsky95}
\Name{Radzihovsky L. \and Toner J.} \REVIEW{Phys. Rev. Lett.}{75}{1995}{4752}.

\bibitem{essafi11}
\Name{Essafi K., Kownacki J.-P.  \and Mouhanna D.} \REVIEW{Phys. Rev.
  Lett.}{106}{2011}{128102}.

\bibitem{alexandre11}
\Name{{Alexandre J.}} \REVIEW{Int. J. Mod. Phys.A}{26}{2011}{4523}.

\bibitem{horava09}
\Name{Horava P.} \REVIEW{Phys. Rev. D}{79}{2009}{084008}.

\bibitem{nakayama11}
\Name{Nakayama Y.} \REVIEW{Gen. Rel. Grav.}{43}{2011}{235}.

\bibitem{henkel02}
\Name{Henkel M.} \REVIEW{Nucl. Phys. B}{641}{2002}{405}.

\bibitem{pleimling01}
\Name{Pleimling M. \and Henkel M.} \REVIEW{Phys. Rev. Lett.}{87}{2001}{125702}.

\bibitem{rutkevich11}
\Name{{Rutkevich S.,  Diehl H. W. \and Shpot M. A.}} \REVIEW{Nucl. Phys.
  B}{843}{2011}{255};    err: ibid {\bf 853}, 210 (2011).

\bibitem{diehl00}
\Name{{Diehl H. W.  \and Shpot M.}} \REVIEW{Phys. Rev. B}{62}{2000}{12338}.

\bibitem{shpot01}
\Name{{Shpot M. A.  \and Diehl H. W.}} \REVIEW{Nucl. Phys. B}{612}{2001}{340}.

\bibitem{shpot08}
\Name{{Shpot M. A., Diehl H. W. \and Pis'mak Yu. M.}} \REVIEW{J. Phys. A:
  Math. Theor.}{41}{2008}{135003}.

\bibitem{diehl01}
\Name{{Diehl H. W.  \and  Shpot M.}} \REVIEW{J. Phys. A}{34}{2001}{9101}.

\bibitem{mergulhao}
\Name{{Mergulh\~{a}o, Jr C. \and  Carneiro C. E. I.}} \REVIEW{Phys. Rev.
  B}{58}{1998}{6047} ibid. {\bf 59} 13954 (1999).

\bibitem{shpot05}
\Name{{ Shpot M. A.,  Pis'mak Yu. M.  \and Diehl H. W.}} \REVIEW{J. Phys. Condens.
  Matter}{17}{2005}{S1947}.
  \bibitem{shpot12}
\Name{{ Shpot M. A.,  Pis'mak Yu. M.}} \REVIEW{Nucl. Phys. B}{862}{2012}{75}.

  \bibitem{grest78}
\Name{{Grest G. S. \and Sak J.}} \REVIEW{Phys. Rev. B}{17}{1978}{3607}.

\bibitem{wetterich93c}
\Name{Wetterich C.} \REVIEW{Phys. Lett. B}{301}{1993}{90}.

\bibitem{bagnuls01}
\Name{Bagnuls C.  \and Bervillier C.} \REVIEW{Phys. Rep.}{348}{2001}{91}.

\bibitem{berges02}
\Name{Berges J., Tetradis N. \and Wetterich C.} \REVIEW{Phys.
  Rep.}{363}{2002}{223}.

  
\bibitem{delamotte03}
\Name{Delamotte B., Mouhanna D. \and Tissier M.} \REVIEW{Phys. Rev.
  B}{69}{2004}{134413}.

\bibitem{pawlowski07}
\Name{Pawlowski J.} \REVIEW{Annals Phys.}{322}{2007}{2831}.

\bibitem{rosten12}
\Name{Rosten O.} \REVIEW{Phys. Rep.}{511}{2012}{177}.

\bibitem{canet03a}
\Name{Canet L., Delamotte B., Mouhanna D. \and Vidal J.} \REVIEW{Phys. Rev.
  D}{67}{2003}{065004}.

\bibitem{canet03b}
\Name{Canet L., Delamotte B., Mouhanna D. \and Vidal J.} \REVIEW{Phys. Rev.
  B}{68}{2003}{064421}.
  
\bibitem{tissier00}
\Name{Tissier M.,  Delamotte B. \and  Mouhanna D.} \REVIEW{Phys. Rev. Lett.}{84}{2000}{5208}.


\bibitem{tissier01b}
\Name{Tissier M., Mouhanna D., Vidal J. \and Delamotte B.} \REVIEW{Phys. Rev.
  B}{65}{2002}{140402}.
  
 \bibitem{kownacki09}
\Name{Kownacki  J.-P.  \and Mouhanna D.} \REVIEW{Phys. Rev.
  E}{79}{2009}{040101}.

\bibitem{bervillier04}
\Name{Bervillier C.} \REVIEW{Phys. Lett. A}{331}{2004}{110}.

\bibitem{hornreich75b}
\Name{Hornreich R. M., Luban M. \and Shtrikman S.} \REVIEW{Phys.
  Lett.}{55A}{1975}{269}.

\bibitem{litim02}
\Name{Litim D. F.} \REVIEW{Nucl. Phys. B}{631}{2002}{128}.

\end{thebibliography}
\end{document}